\documentclass{appolb}
\usepackage{epsfig}
\usepackage{cite}
\usepackage{amsmath}
\usepackage{psfrag}
\usepackage{color}

\def\theequation{\thesection.\arabic{equation}}

\newcommand{\nn}{\nonumber}

\newcommand{\eps}{\varepsilon}
\newcommand{\RE}{\text{Re}}
\newcommand{\IM}{\text{Im}} 

\newcommand{\newsection}[1]{\section{#1}\setcounter{equation}{0}}
\newcommand{\bea}{\begin{eqnarray}}
\newcommand{\eea}{\end{eqnarray}}

\newcommand{\be}{\begin{equation}}
\newcommand{\ee}{\end{equation}}
\newcommand{\bi}{\begin{itemize}}
\newcommand{\ei}{\end{itemize}}
\newcommand{\ord}{{\cal O}}

\def\kpn{K^+\rightarrow\pi^+\nu\bar\nu}
\def\klpn{K_{L}\rightarrow\pi^0\nu\bar\nu}

\newcommand{\tev}{\, {\rm TeV}}
\newcommand{\gev}{\, {\rm GeV}}

\def\simge{\mathrel{\rlap{\raise 0.511ex \hbox{$>$}}{\lower 0.511ex \hbox{$\sim$}}}}
\def\simle{\mathrel{\rlap{\raise 0.511ex \hbox{$<$}}{\lower 0.511ex \hbox{$\sim$}}}}

\headtitle{A Guide to FCNCs in the Littlest Higgs Model with T-Parity}
\headauthor{Monika Blanke and Andrzej J. Buras}
\begin{document}

\eqsec
\title{A GUIDE TO\\ FLAVOUR CHANGING NEUTRAL CURRENTS \\IN THE LITTLEST HIGGS MODEL WITH T-PARITY}
\author{Monika Blanke$^{a,b}$ and Andrzej J. Buras$^a$
\address{$^a$Technische Universit{\"a}t M{\"u}nchen,
  Physik Department,\\ D-85748 Garching, Germany\\
$^b$Max-Planck-Institut f{\"u}r Physik (Werner-Heisenberg-Institut), \\
D-80805 M{\"u}nchen, Germany}}
\maketitle

%\centerline{\bf Abstract}
\begin{abstract}
\noindent
Flavour changing neutral current processes, being strongly suppressed in the Standard Model (SM), provide a unique window to new physics at scales much above the electroweak scale. Here, we summarize the recent progress in flavour physics studies of the Littlest Higgs model with T-parity, both in the quark and lepton sector. Particular emphasis is put on various correlations that could distinguish this model from other extensions of the SM.
\end{abstract}

%%% MAIN TEXT

\newsection{Introduction}

Until now, essentially all available data have shown an impressive agreement with the Standard Model (SM) predictions. In particular, electroweak precision tests and constraints on {flavour changing neutral current (FCNC) processes} put very stringent constraints on physics {beyond the SM, requiring it to appear first at scales $\ord(10\tev)$. On the other hand, new physics (NP) is expected already at scales $\ord(1\tev)$} in order to offer a natural explanation to the smallness of the Higgs mass.

In the case of flavour physics, the simplest solution to this so-called \emph{little hierarchy problem} is provided by the concept of Minimal Flavour Violation (MFV) \cite{mfv1,mfv2,mfvold}, in which no new sources of flavour and CP violation beyond the SM CKM matrix \cite{ckm} are present. While this approach is clearly a very elegant way to account for small NP effects in flavour violating observables, there is still room left for departures from the MFV framework, in particular in observables that have not been measured so far, such as CP violation in the $B_s$ meson system and some rare $K$ and $B$ decays. Moreover, as we will see below, very large departures from the SM expectations are still possible in lepton flavour violating (LFV) decays.

\newsection{The Littlest Higgs Model with T-Parity}

While Supersymmetry is so far the leading candidate for NP beyond the SM, offering a solution to the little hierarchy problem and allowing at the same time for large effects in FCNC observables, other alternatives have been developed over the past years. Among those, Little Higgs models \cite{lhold,lhreview} are one of the most popular possibilities. Here, the Higgs boson is interpreted as pseudo-Goldstone boson of a spontaneously broken global symmetry, and is thus massless at tree level. As the global symmetry is also broken explicitly by gauge and Yukawa interactions, a Higgs potential is generated radiatively.

In order to then prevent the Higgs squared mass from dangerous quadratically divergent contributions at one loop level, the \emph{collective symmetry breaking} mechanism is introduced. All couplings explicitly breaking the global symmetry are introduced in such a way that, as long as only one coupling is present, still enough of the global symmetry is preserved to protect the Higgs mass. Only when more than one coupling is non-zero, the symmetry is broken and corrections to the Higgs mass arise. These corrections, however, turn out to be at most logarithmically divergent at one loop, and therefore safely small.

As the mechanism of this spontaneous symmetry breaking is not specified, but merely described by a non-linear sigma model, Little Higgs models are effective theories with an {ultraviolet (UV)} cutoff $\Lambda\sim 10\tev$. Therefore, at a certain level of accuracy one has to worry about effects coming from the UV completion of the model. We will return to this issue in Section \ref{sec:rare}, where we discuss the interpretation of left-over logarithmic divergences that appear in the calculation of rare decay branching ratios.

The most economical, in matter content, Little Higgs model is the Littlest Higgs model (LH) \cite{l2h}, which has been studied extensively in the recent literature. In this model, which is based on an $SU(5)/SO(5)$ non-linear sigma model, new heavy gauge bosons $(W_H^\pm,Z_H,A_H)$, a heavy vectorlike quark $(T)$ and a heavy scalar triplet $(\Phi)$ are present. Due to tree level contributions of the new particles and to the breakdown of the custodial $SU(2)$ symmetry, {the NP scale $f$ is required} to be at least $2-3\tev$ in order to be consistent with electroweak precision constraints \cite{lhewpt}. {As a consequence of this, and because the LH model belongs to the class of models with constrained MFV (CMFV) \cite{mfv1,BBGT}, where the departures from the SM are required to be small \cite{Bobeth}, the NP effects in FCNC observables turn out to be at most 20\% \cite{lh-FCNC,BPUB}. These are in principle good news for this model, but its flavour phenomenology is not very exciting.}

The situation improves significantly once an additional discrete symmetry, T-parity, is introduced \cite{tparity}. Under this symmetry, all SM particles and the heavy $T\equiv T_+$ are T-even, whereas the new heavy gauge bosons $(W_H^\pm,Z_H,A_H)$ and the scalar triplet $(\Phi)$ are T-odd. Consequently the new particles can contribute to electroweak precision observables only at the loop level, thus making it possible to lower the scale $f$ down to $\simge 500\gev$ without violating existing constraints \cite{LHT-EWPT}.

A consistent implementation of T-parity in the fermion sector makes also the introduction of three doublets of mirror quarks ($q_H^i$) and three doublets of mirror leptons ($\ell_H^i$), and of a T-odd singlet quark ($T_-$), necessary \cite{mirror}. The mirror fermions have new flavour violating interactions with the SM quarks, mediated by the heavy gauge bosons, which are parameterized by 4 new CKM-like mixing matrices: $V_{Hu}$ and $V_{Hd}$ in the quark sector, and $V_{H\nu}$ and $V_{H\ell}$ in the lepton sector \cite{Hubisz}.
The indices of the new mixing matrices denote which of the SM fermions is involved in the interaction. Therefore, $V_{Hd}$, the most important mixing matrix for the study of FCNCs in $K$ and $B$ meson systems, parameterizes interactions of SM down-type quarks with mirror quarks. {Similarly, $V_{H\ell}$ parameterizes interactions of SM charged leptons with mirror leptons, being thus relevant for the study of charged LFV processes.}

In contrast to the CKM matrix, from which five phases can be rotated away due to phase redefinitions of the SM quark fields, the new mixing matrix $V_{Hd}$ can be parameterized by three mixing angles and \emph{three} complex phases \cite{SHORT}, as only the three mirror quark doublets can be used to absorb unphysical phases. Therefore, $V_{Hd}$ can be parameterized as follows \cite{SHORT}:
\bea
V_{Hd}&=& \begin{pmatrix}
1 & 0 & 0\\
0 & c_{23}^d & s_{23}^d e^{- i\delta^d_{23}}\\
0 & -s_{23}^d e^{i\delta^d_{23}} & c_{23}^d\\
\end{pmatrix}\cdot
 \begin{pmatrix}
c_{13}^d & 0 & s_{13}^d e^{- i\delta^d_{13}}\\
0 & 1 & 0\\
-s_{13}^d e^{ i\delta^d_{13}} & 0 & c_{13}^d\\
\end{pmatrix}\nn\\
&&\cdot
 \begin{pmatrix}
c_{12}^d & s_{12}^d e^{- i\delta^d_{12}} & 0\\
-s_{12}^d e^{i\delta^d_{12}} & c_{12}^d & 0\\
0 & 0 & 1\\
\end{pmatrix}.
\eea
In complete analogy, $V_{H\ell}$ can be parameterized by three mixing angles and three (non-Majorana) phases. The matrices $V_{Hu}$ and $V_{H\nu}$ are then determined through
\be
V_{Hu}=V_{Hd}V_\text{CKM}^\dagger\,,\qquad
V_{H\nu}=V_{H\ell}V_\text{PMNS}\,,
\ee
where the hermitian conjugate on the r.\,h.\,s. of the first, but not the second, equation is due to the fact that the PMNS matrix is defined through neutrino mixing, while the CKM matrix describes mixing in the down-type quark sector.

For the study of FCNC processes in the quark sector it is useful to define  within the SM the factors ${(i=u,c,t)}$
\be\label{eq:lambda_i}
\lambda_i^{(K)}=V^{*}_{is}V^{}_{id}\,,\qquad
\lambda_i^{(d)}=V^{*}_{ib}V^{}_{id}\,,\qquad
\lambda_i^{(s)}=V^{*}_{ib}V^{}_{is}\,,
\ee
that govern $K$, $B_d$ and $B_s$ meson systems, respectively. In the T-odd sector of the LHT model the same role is played by {($i=1,2,3$ corresponding to the three mirror quark generations with masses $m_{Hi}$)} 
\cite{BBPTUW,BBPRTUW}
\be
\xi_i^{(K)}=V^{*is}_{Hd}V^{id}_{Hd}\,,\qquad
\xi_i^{(d)}=V^{*ib}_{Hd}V^{id}_{Hd}\,,\qquad
\xi_i^{(s)}=V^{*ib}_{Hd}V^{is}_{Hd}\,.
\ee
Similarly, for the lepton sector, one has \cite{BBDPT}
\begin{equation}\label{eq:chi}
\chi_i^{(\mu e)}=V^{*ie}_{H\ell}V^{i\mu}_{H\ell}\,,\qquad
\chi_i^{(\tau e)}=V^{*ie}_{H\ell}V^{i\tau}_{H\ell}\,,\qquad
\chi_i^{(\tau\mu)}=V^{*i\mu}_{H\ell}V^{i\tau}_{H\ell}\,,
\end{equation}
entering $\mu\to e$, $\tau\to e$ and $\tau\to\mu$ transitions, respectively.

If the new particles are discovered and their masses determined at the LHC, the only free parameters of the model will be the parameters of the new mixing matrices $V_{Hd}$ and $V_{H\ell}$, which can in principle be determined from quark and lepton flavour violating processes.

\newsection{Flavour Changing Neutral Currents in the Quark Sector}

\subsection{Hints for New Physics beyond the SM and MFV}

The value of the mass difference $\Delta M_s$, having recently been measured \cite{CDF,D0} to be
\be
\Delta M_s = (17.77\pm0.10\pm0.07)\,\text{ps}^{-1}\,,
\ee
turned out to be surprisingly below the SM expectation, albeit still within the theoretical uncertainties. On the other hand, in \cite{BBbound} it has been shown that in CMFV models $\Delta M_s \ge (\Delta M_s)_\text{SM}$ is predicted. In order to suppress $\Delta M_s$ below its SM value, new sources of flavour violation, new relevant operators or contributions from {Majorana fermions or new heavy $U(1)$ gauge bosons in box diagrams}  are required.

\begin{figure}
\center{\epsfig{file=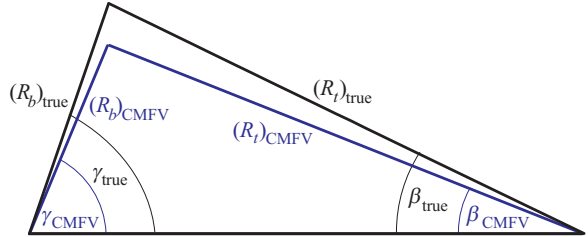,scale=0.7}}
\caption{RUT and UUT, constructed from the central values for $\gamma$, $|V_{ub}/V_{cb}|$, $\Delta M_d/\Delta M_s$ and $S_{\psi K_S}$ \cite{BBGT}. \label{fig:rut-uut}}
\end{figure}

Another hint for new sources of flavour violation is found when comparing the \emph{reference unitarity triangle} (RUT) \cite{RUT}, constructed from the tree level values of $\gamma$ and $|V_{ub}/V_{cb}|$ with the \emph{universal unitarity triangle} (UUT) \cite{mfv1}, defined through $\Delta M_d/\Delta M_s$ and $S_{\psi K_S}$ (see Fig.~\ref{fig:rut-uut} and \cite{BBGT}). While the former is to an excellent accuracy independent of any NP contribution, the latter is universal only within the CMFV framework. It turns out that there is a $2.3\sigma$ discrepancy between the ``true'' value of $\beta$, determined from the RUT, and its CMFV value, determined from $S_{\psi K_S}$. Provided the large value of $(R_b)_\text{true}\propto |V_{ub}|$, coming from inclusive semileptonic decays, will be confirmed by more accurate experimental analyses,  such a discrepancy could only be resolved by the presence of a new CP-violating phase $\varphi_{B_d}\approx-5^\circ$ and would be a clear signal of physics beyond the MFV scenario. {Similar conclusions have been reached for instance in \cite{beta-problem}.}

\subsection{$K$ and $B$ Physics in the LHT Model}

The first study of particle-antiparticle mixing in the LHT model has been presented in \cite{Hubisz}, where the mass differences $\Delta M_K$, $\Delta M_d$, $\Delta M_s$ and $\Delta M_D$ and the CP-violating parameter $\eps_K$ have been considered. In \cite{BBPTUW} this analysis has been extended to the theoretically cleaner observables $S_{\psi K_S}$, $S_{\psi\phi}$, $A^{d,s}_\text{SL}$ and $\Delta\Gamma_{d,s}$, which allowed a deeper insight into the flavour structure of the LHT model. In the latter paper also the decay $B\to X_s\gamma$ has been considered. 

The study of quark FCNC processes in the LHT model has then been completed by the analysis of \cite{BBPRTUW}, where most prominent rare and CP-violating $K$ and $B$ decays have been analyzed in detail. More precisely the decays $K\to\pi\nu\bar\nu$,  $K_L\to\pi^0\ell^+\ell^-$, $B\to X_s\ell^+\ell^-$, $B_{d,s}\to\mu^+\mu^-$, $B\to X_{d,s}\nu\bar\nu$ and $B\to\pi K$ have been considered and correlations between various observables have been studied.

{Very recently, the rare top quark decays $t\to cV$ ($V=\gamma,Z,g$) have been analyzed in \cite{Hong-Sheng}.}

\subsubsection{Benchmark Scenarios}

In order to gain a global view about possible signatures of mirror fermions in the processes considered, several benchmark scenarios for the structure of the $V_{Hd}$ matrix and the mirror fermion spectrum have been introduced in \cite{BBPTUW,BBPRTUW}. The following two turned out to be most interesting, as they allow for large effects in the $B_s$ and $K$ systems, respectively, while being consistent with all available constraints:

{\bf \boldmath${B_s}$\unboldmath~scenario:} Here, in order to suppress the new contributions to $\Delta M_K$ and $\eps_K$, the first two mirror quark generations are chosen to be quasi-degenerate. Furthermore, an inverted hierarchy for the $V_{Hd}$ matrix,
\be
s_{23}^d \ll s_{13}^d\le s_{12}^d\,,
\ee 
with respect to the usual CKM one,
\be
s_{13} \ll s_{23} \ll s_{12}\,,
\ee
 is chosen. Like this, large effects in CP-violating observables in the $B_s$ system are possible.

{\bf \boldmath${K}$\unboldmath~scenario:} Again, the first two generations of mirror quarks are quasi-degenerate. Moreover, the $V_{Hd}$ parameters satisfy
\be
c_{12}^d=s_{12}^d=\frac{1}{\sqrt{2}}\,,\qquad s_{23}^d=\frac{s_{13}^d}{\sqrt{1+{s_{13}^d}^2}}\,.
\ee
This structure leads to possible large enhancements of rare $K$ decay branching ratios.

In addition, in order not to miss any interesting effect, a general scan over the whole parameter space of the LHT model has been performed.

\subsubsection{Particle-Antiparticle Mixing, CP Violation and $B\to X_s\gamma$}\label{sec:LHT1}

Here we summarize the main findings of the phenomenological analysis of particle-antiparticle mixing, CP violation and the decay $B\to X_s\gamma$ in the LHT model presented in \cite{BBPTUW}.

The LHT model can be made consistent with all available data on FCNC processes, provided the weak mixing matrix 
$V_{Hd}$ exhibits a hierarchical structure and the mass spectrum of 
mirror fermions is quasi-degenerate. However, the structure of the mixing matrix 
$V_{Hd}$ can differ 
significantly from the known structure of the CKM matrix so that 
interesting departures from MFV correlations between various 
processes are possible. We will return to this issue in Section \ref{sec:rare}.

The T-even sector of the LHT model, that represents the only LHT contribution to FCNC processes in the CMFV limit of exactly degenerate mirror fermions,
is not favoured by the data as independently of the parameters of this sector 
$\Delta M_s > (\Delta M_s)_{\rm SM}$ and the possible discrepancy between 
the value of the CP asymmetry $S_{\psi K_S}$ and large 
values of $|V_{ub}|$ cannot be removed.

Using the full structure of new flavour and CP-violating interactions 
encoded in $V_{Hd}\not=V_{\rm CKM}$, regions in the parameter
space of the LHT model have been identified in which possible problems of the SM can be 
cured, large CP-violating effects in the $B_s$ system are predicted
 and the mass difference 
$\Delta M_s$ is found to be smaller than $(\Delta M_s)_{\rm SM}$ as
{possibly hinted} by the recent result of the CDF collaboration \cite{CDF}.

\begin{figure}
\begin{minipage}{6.5cm}
a)\vspace{-.3cm}
\center{\epsfig{file=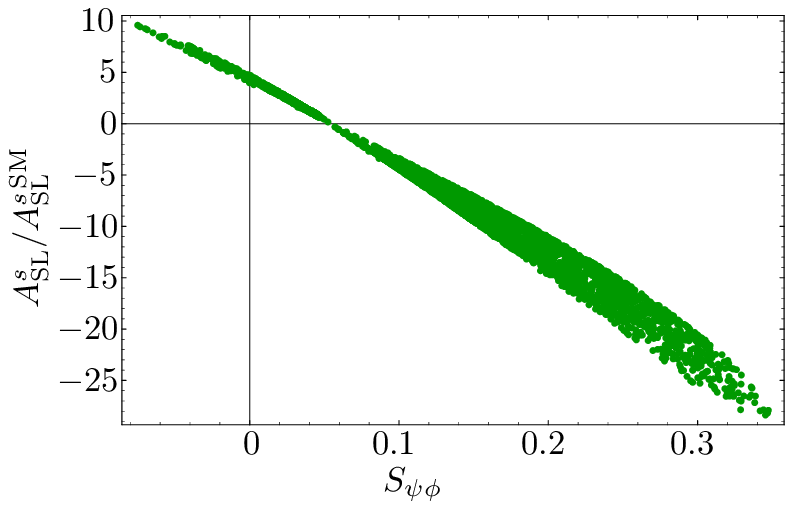,scale=.7}}
\end{minipage}
\begin{minipage}{6cm}
b)\vspace{-.6cm}
\center{\epsfig{file=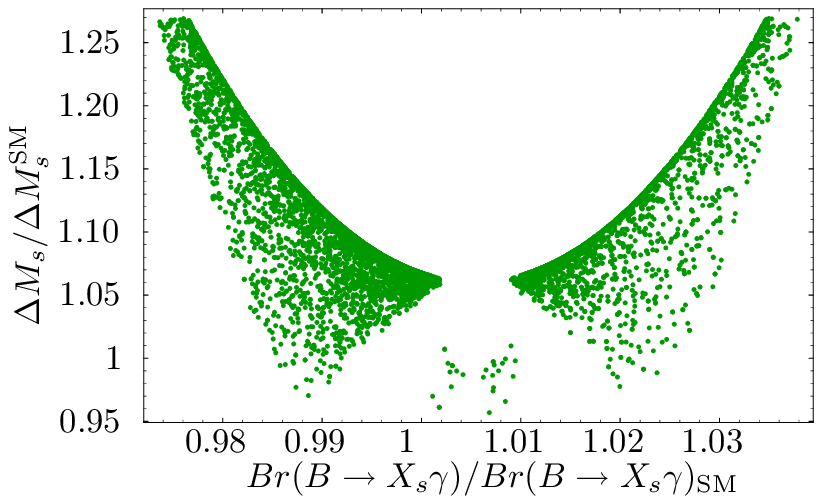,scale=.7}}
\end{minipage}
\caption{a) Correlation between $A^s_\text{SL}/(A^s_\text{SL})_\text{SM}$ and $S_{\psi\phi}$ in the ``$B_s$ scenario'' \cite{BBPTUW}.  b)~Correlation between $\Delta M_s$ and $Br(B\to X_s\gamma)$, normalized to their SM values, in the ``$B_s$ scenario'' \cite{BBPTUW}. \label{fig:LHT1}}
\end{figure}

In particular, as shown in Fig.~\ref{fig:LHT1}.a), in the ``$B_s$ scenario''
significant enhancements of the CP asymmetries 
$S_{\psi \phi}$ and $A_{\rm SL}^s$ relative to the SM
are possible, while satisfying all existing 
constraints. We emphasize that the correlation in Fig.~\ref{fig:LHT1}.a) is much stronger than in the model independent analysis of \cite{Ligeti}.

On the other hand, the effects from mirror fermions in the
 $B\to X_s\gamma$ decay, shown in Fig.~\ref{fig:LHT1}.b), turn out to be smaller than in the $\Delta B=2$ 
 transitions, which should be welcomed as the SM is here in a 
rather 
good shape, even if the most recent NNLO values are slightly below the data \cite{bsgNNLO}. Typically the LHT effects are below $4\%$, and also the effects on the corresponding CP asymmetry are small.

\subsubsection{Rare and CP-violating $K$ and $B$ Decays}\label{sec:rare}

Let us now turn to rare and CP-violating $K$ and $B$ decays in the LHT model, which have been discussed extensively in \cite{BBPRTUW}.

Before stating the results of that analysis, it is illuminating to first have a look at the general structure of weak decay amplitudes in the LHT model. In the SM, the general structure of a decay amplitude is given by
\be
\mathcal{A}_\text{SM}=\sum_i B_i^\text{SM}\eta_i^\text{QCD} \lambda_i^{(q)} F_\text{SM}(m_i,M_W)\,,
\ee
where $B_i^\text{SM}$ and $\eta_i^\text{QCD}$ are non-perturbative factors and QCD corrections, respectively. $\lambda_i^{(q)}$ ($q=K,d,s$) denote the relevant CKM factors in \eqref{eq:lambda_i} and $F_\text{SM}$ the SM loop functions. This structure is modified in the LHT model as follows:
\be
\mathcal{A}_\text{LHT}=\sum_i B_i^\text{SM}\eta_i^\text{QCD} \Big[\lambda_i^{(q)} F_\text{even}(m_i,M_W,f,x_L)+ \xi_i^{(q)} \,F_\text{odd}(m_{Hi},f) \Big]\,,
\ee
{where the parameter $x_L$ describes the mixing in the T-even top sector. The first term} corresponds to the T-even contributions and the second term describes the mirror fermion contributions. Obviously the latter contributions constitute a new source of flavour and CP violation that can lead to large NP effects. It is important to note that in the LHT model no new operators appear beyond those that are present already in the SM; therefore the non-perturbative uncertainties remain the same as in the SM and NP effects are described entirely by short distance physics. {As generally top quark effects are dominant in the SM, it is useful to encode all NP effects into the loop functions $F^q_\text{LHT}$ multiplying $\lambda_t^{(q)}$, {which can then be written as}
\be
F^q_\text{LHT}(m_t,m_{Hi},M_W,f,x_L) = F_\text{even}(m_t,M_W,f,x_L)+ \sum_i\frac{\xi_i^{(q)}}{\lambda_t^{(q)}} \,F_\text{odd}(m_{Hi},f)\,.
\ee}
 Bearing in mind the hierarchy of $\lambda_t^{(q)}$ in the SM, i.\,e. $\lambda_t^{(K)}\simeq 4\cdot 10^{-4}$, $\lambda_t^{(d)}\simeq 1\cdot10^{-2}$ and $\lambda_t^{(s)}\simeq 4\cdot10^{-2}$, one can roughly estimate the relative size of the NP contributions to rare $K$, $B_d$ and $B_s$ decays: Largest effects are to be expected in rare $K$ decays, while the NP contributions are smaller by more than an order of magnitude in the $B_d$ and even by two orders of magnitude in the $B_s$ system. We will return to the consequences of this obvious breakdown of universality later on.

From the technical side, a left-over logarithmic divergence  appears in the calculation of rare decay branching ratios in the LHT model. Such a divergence has already been found and discussed in \cite{BPUB} in the context of rare decays in the LH model without T-parity. In \cite{BBPRTUW} it has been discovered that imposing T-parity removes such divergences from the T-even sector, while they now arise in the T-odd sector of the model. An explicit calculation in the 't~Hooft-Feynman gauge shows that this divergence, $\delta_\text{div}$, arises from a single diagram and follows entirely from the interactions of the
Goldstone bosons of the dynamically broken global $SU(5)$ symmetry
with the light fermions (being mediated by $W_H^\pm$). Furthermore $\delta_\text{div}$ is flavour universal and can therefore be traded for one observable, which can be determined from experiment once more data on FCNC processes are available. On the other hand, if the {UV completion} of the model were known, $\delta_\text{div}$ would be replaced by some cutoff independent expression through matching of the full theory to the effective theory. Meanwhile, $\delta_\text{div}$ can be estimated as
\be
\delta_\text{div} =\frac{1}{\eps}+\log\frac{\mu^2}{M_{W_H}^2}\longrightarrow \log\frac{\Lambda^2}{M_{W_H}^2}\,,
\ee
where $\Lambda = 4\pi f \sim 10\tev$ is the {UV cutoff} of the LHT model. In Table \ref{tab:classification} we list which of the observables considered in \cite{BBPTUW,BBPRTUW,BBDPT} suffer from the presence of this UV divergence (class A) and which do not (class B). Certainly the predictions for class B are more reliable, but we believe that also the above estimate of the divergent contribution should give at least qualitatively correct predictions for class A observables.   

\begin{table}
\begin{center}
\begin{tabular}{|l|c|c|}
\hline
 & class A & class B \\\hline
quark sector & $K\to\pi\nu\bar\nu,K_L\to\pi^0\ell^+\ell^-,$& $\Delta M_K,\eps_K,\Delta M_{d,s},\Delta M_D,$ \\
 & $B\to X_s\ell^+\ell^-,B_{d,s}\to\mu^+\mu^-,$ & $S_{\psi K_S},S_{\psi\phi},A^{d,s}_\text{SL},\Delta\Gamma_{d,s},$\\
 & $B\to X_{d,s}\nu\bar\nu$ & $B\to X_{d,s}\gamma$\\\hline
lepton sector & $\ell_i^-\to\ell_j^-\ell_j^+\ell_j^-,\tau^-\to\ell_i^-\ell_k^+\ell_k^-,$ & $\ell_i\to\ell_j\gamma,\tau^-\to\ell_i^-\ell_k^+\ell_i^-,$\\
& $\tau\to\ell\pi,\ell\eta,\ell\eta',$ & $K_{L,S}\to\mu e,K_{L,S}\to\pi^0\mu e,$ \\
 & $\mu-e$ conversion & $B_{d,s}\to\ell_i\ell_k,(g-2)_\mu$ \\\hline
\end{tabular}
\end{center}
\caption{Classification of observables and decays depending on their sensitivity to the UV completion. Decays of class A bear some sensitivity to the physics above the cutoff, reflected through the left-over logarithmic divergence, while decays of class B are free from this divergence \cite{BBPRTUW,BBDPT}.\label{tab:classification}}
\end{table}

Now we are prepared to review the main results of the phenomenological analysis performed in \cite{BBPRTUW}.

{The most evident departures from the SM predictions are found for
    CP-violating observables that are strongly suppressed within that
    model. These are the branching ratio for $K_L \to \pi^0 \nu \bar \nu$ and, as already discussed in Section \ref{sec:LHT1},
    the {CP asymmetry} $S_{\psi \phi}$. As seen in  Fig.~\ref{fig:KLKp}, there exist two possible branches for $Br(\klpn)$ and $Br(\kpn)$. The first one implies a simultaneous enhancement of these two decays by {at most a factor 10 and 5}, respectively, while on the second one, $\klpn$ stays SM-like and $\kpn$ can be enhanced by an order of magnitude.

\begin{figure}
\center{\epsfig{file=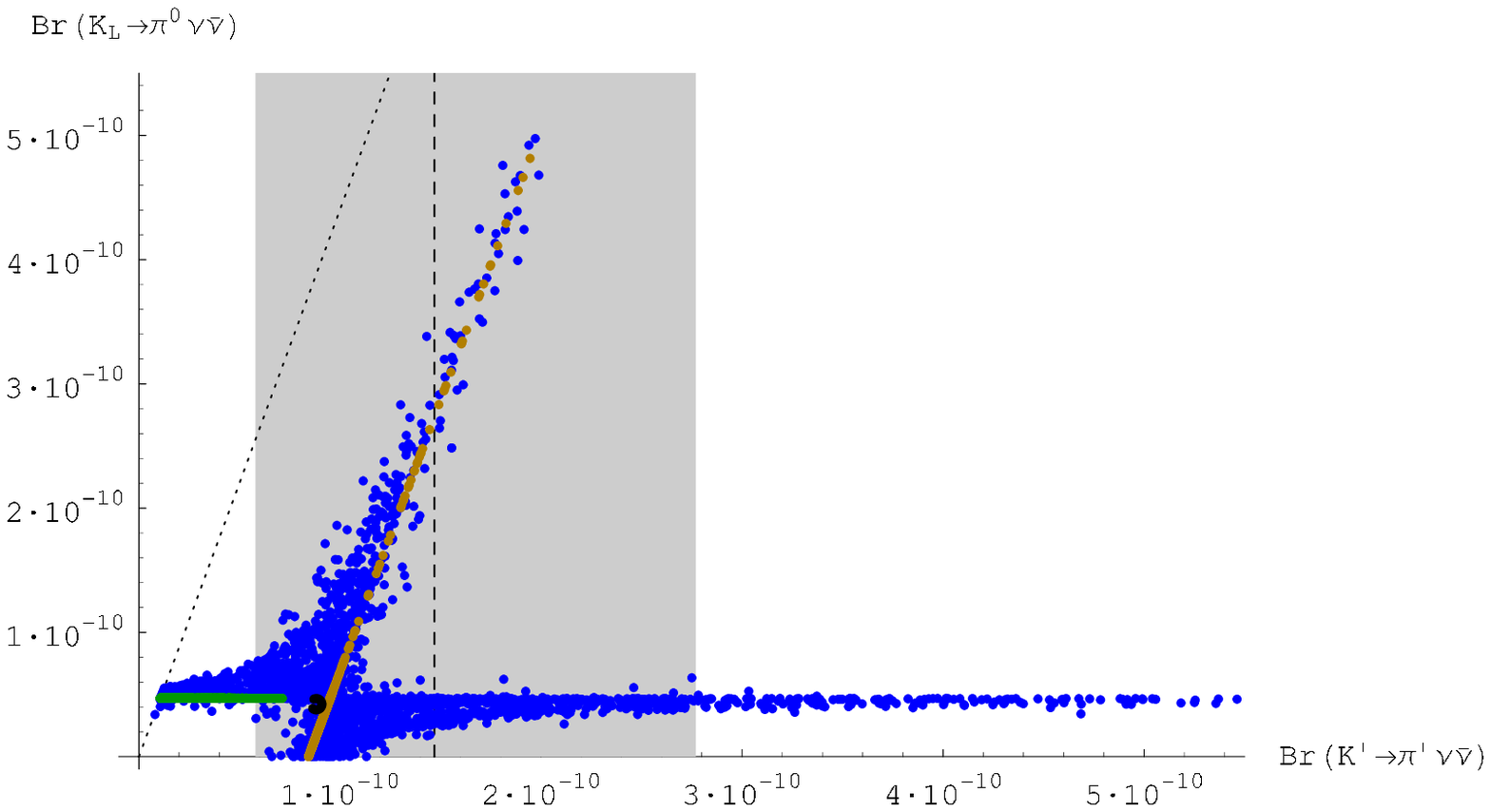,scale=0.5}}
\vspace{.1cm}
\caption{$Br(K_L\to\pi^0\nu\bar\nu)$ as a function of $Br(K^+\to\pi^+\nu\bar\nu)$ in the scenarios considered \cite{BBPRTUW}. The shaded area displays the experimental $1\sigma$-range for $Br(K^+\to\pi^+\nu\bar\nu)$, while the dotted line represents the Grossman-Nir bound \cite{GNbound}. The SM values are indicated by the black point.\label{fig:KLKp}}
\end{figure}

\begin{figure}
\begin{minipage}{6.5cm}
a)\vspace{-.6cm}
\center{\epsfig{file=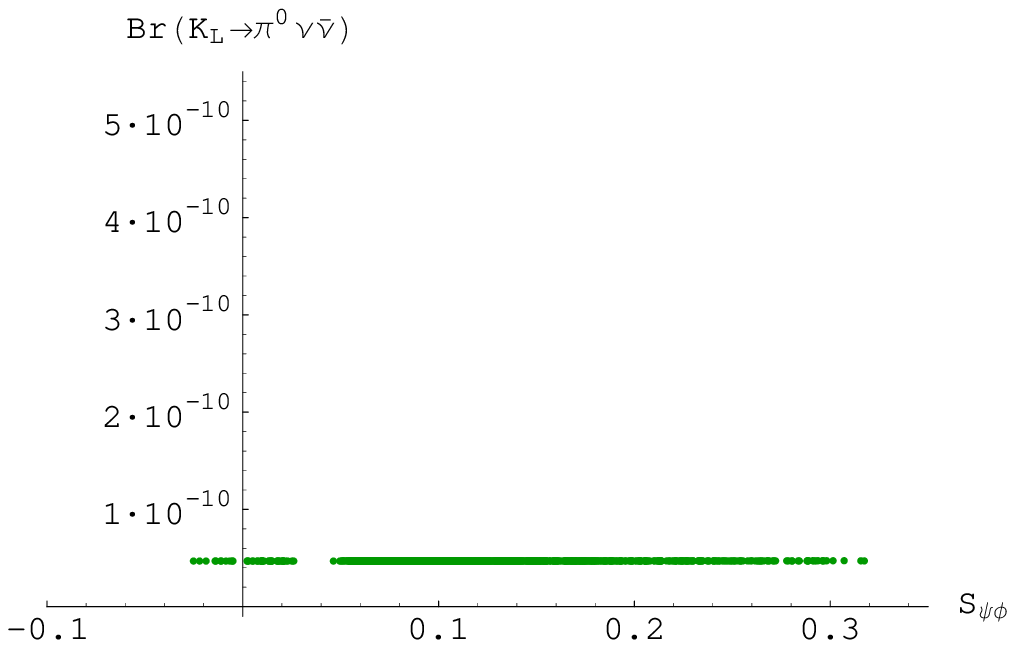,scale=.5}}
\end{minipage}
\begin{minipage}{6cm}
b)\vspace{-.6cm}
\center{\epsfig{file=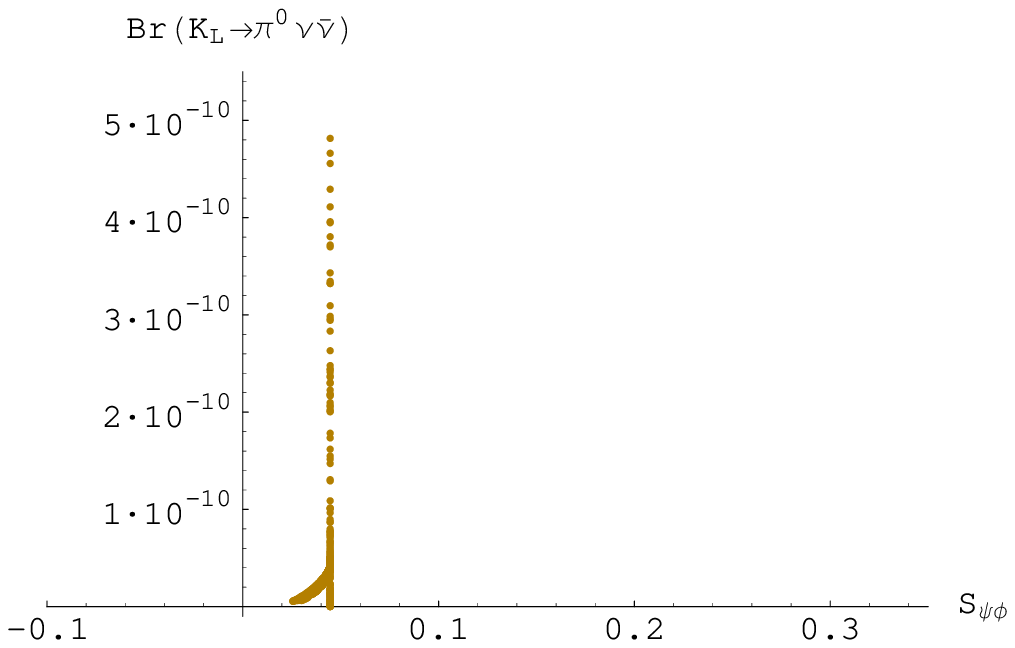,scale=.5}}
\end{minipage}
\center{
\begin{minipage}{6cm}
c)\vspace{-.6cm}
\center{\epsfig{file=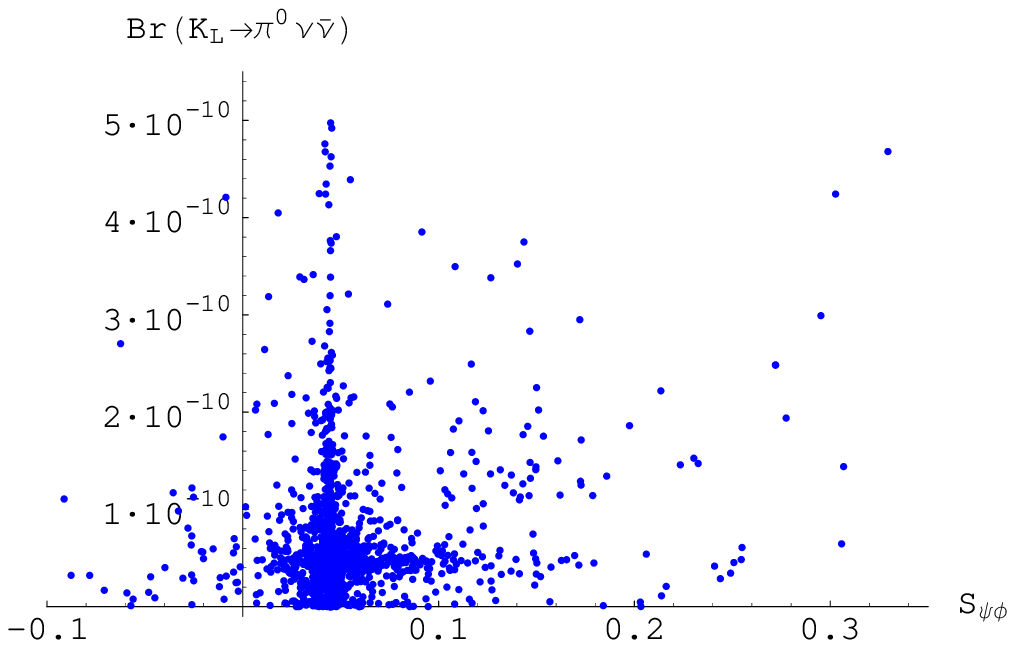,scale=.5}}
\end{minipage}}
\vspace{.1cm}
\caption{Correlation between $Br(K_L\to\pi^0\nu\bar\nu)$ and $S_{\psi\phi}$ in the ``$B_s$ scenario'' (a), in the ``$K$ scenario'' (b) and in the general scan (c) \cite{BBPRTUW}.\label{fig:KL-Spsiphi}}
\end{figure}

Next, in Fig.~\ref{fig:KL-Spsiphi} we show the correlation between $Br(\klpn)$ and $S_{\psi\phi}$. While in the ``$K$'' and ``$B_s$'' scenarios considered, only one of these two quantities can differ significantly from its SM expectation, in the general scan some fine-tuned regions of the parameter space appear, where both observables can be enhanced by a factor 10.

\begin{figure}
\begin{minipage}{6.3cm}
a)\vspace{-.2cm}
\center{\epsfig{file=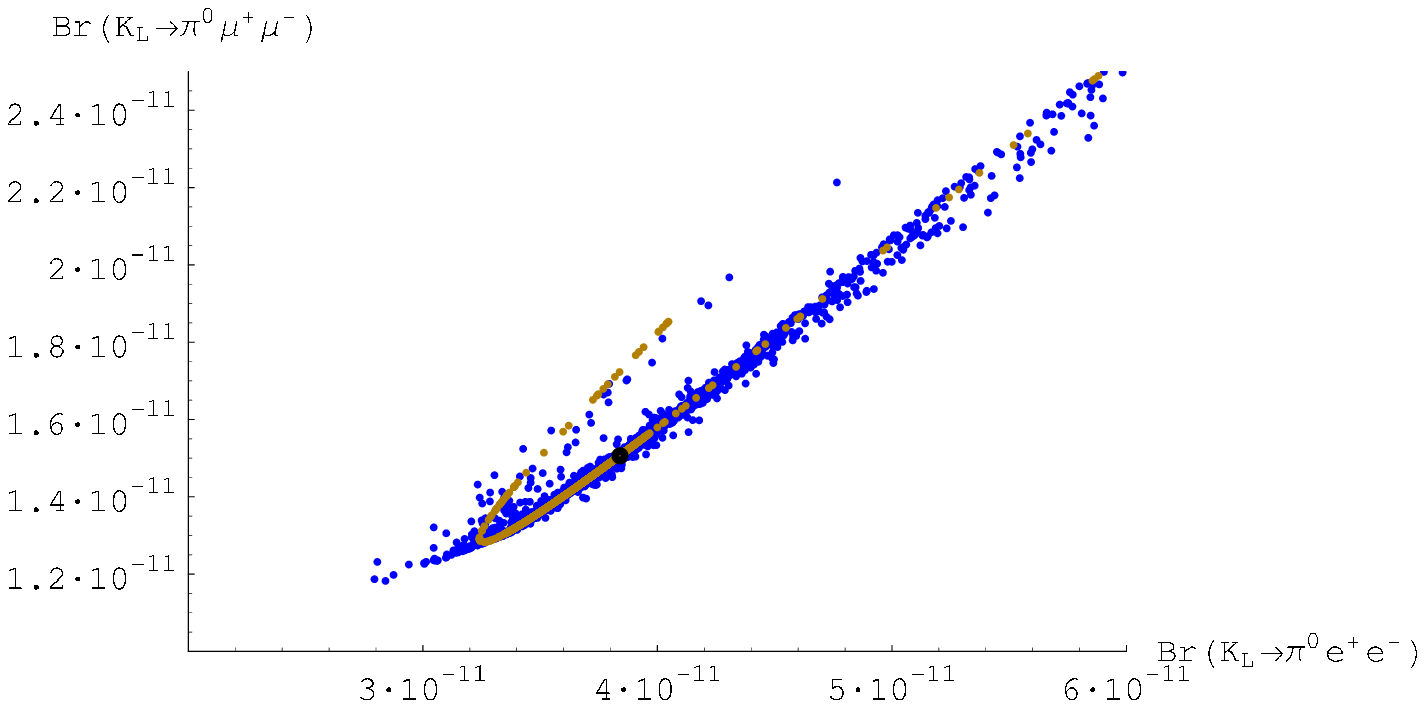,scale=.4}}
\end{minipage}
\begin{minipage}{6.4cm}
b)\vspace{-.2cm}
\center{\epsfig{file=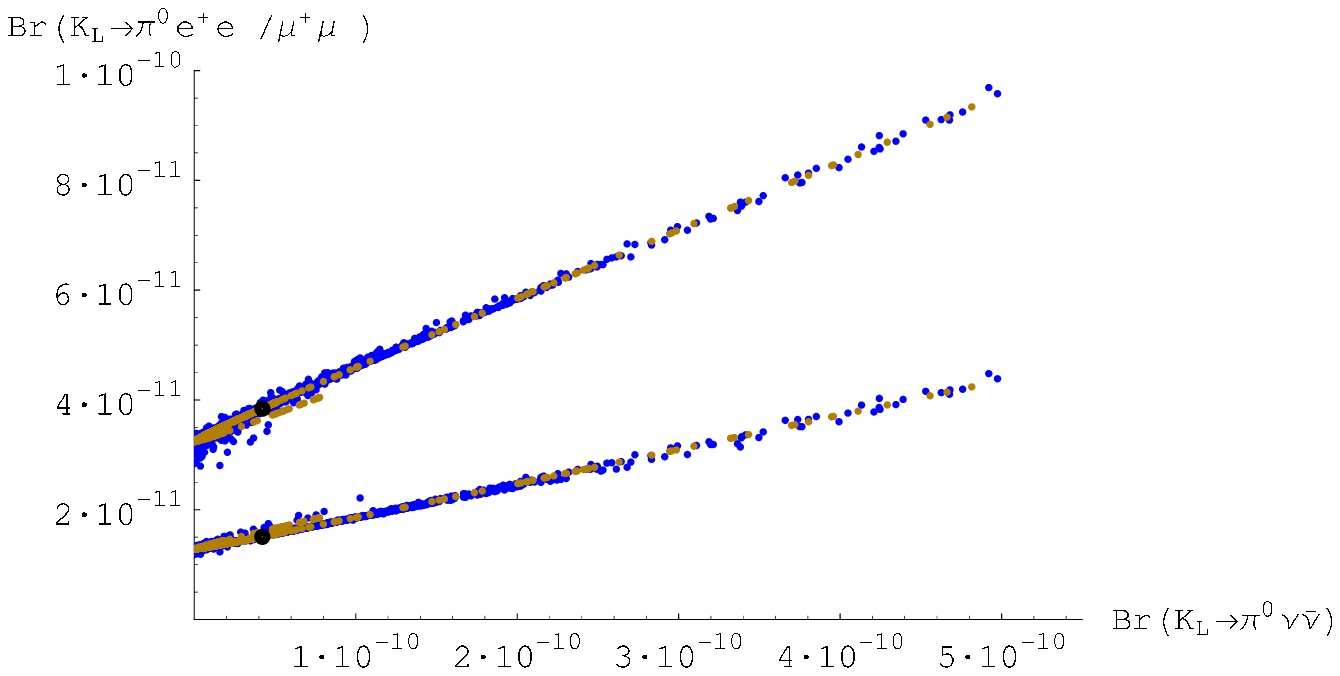,scale=.42}}
\end{minipage}
\vspace{.1cm}
\caption{a) Correlation between $Br(K_L\to\pi^0e^+e^-)$ and $Br(K_L\to\pi^0\mu^+\mu^-)$ \cite{BBPRTUW}. b) $Br(K_L\to\pi^0e^+e^-)$ (upper curve) and $Br(K_L\to\pi^0\mu^+\mu^-)$ (lower curve) as functions of $Br(K_L\to\pi^0\nu\bar\nu)$ \cite{BBPRTUW}. The SM values are indicated by the black points.\label{fig:KLllKLnn}}
\end{figure}

{Large departures from the SM expectations are also possible for the decays $K_L \to
  \pi^0 \ell^+ \ell^-$, which can be enhanced by roughly a factor 2 and are found to be strongly correlated, as shown in Fig.~\ref{fig:KLllKLnn}.a). {Indeed, in the LHT model this correlation turns out to be much stronger than in the MSSM \cite{KLll}.} Furthermore, as seen in Fig.~\ref{fig:KLllKLnn}.b), a strong correlation between $Br(K_L \to
  \pi^0 \ell^+ \ell^-)$ and $Br(\klpn)$ exists, which we expect to be valid also in other NP scenarios, at least if no new operators are relevant.}

{The branching ratios for $B_{s,d} \to \mu^+ \mu^-$ and $B \to X_{s,d}
    \nu \bar \nu$, instead,  are modified by at most $50\%$ and $35\%$,
    respectively, and the effects of new electroweak penguins in $B \to \pi K$
    are small, in agreement with the recent data. Also the effects in $B\to X_{s,d}\ell^+\ell^-$ turn out to be small and therefore in good agreement with the data.}

\begin{figure}
\begin{minipage}{6.4cm}
a)\vspace{-.1cm}
\center{\epsfig{file=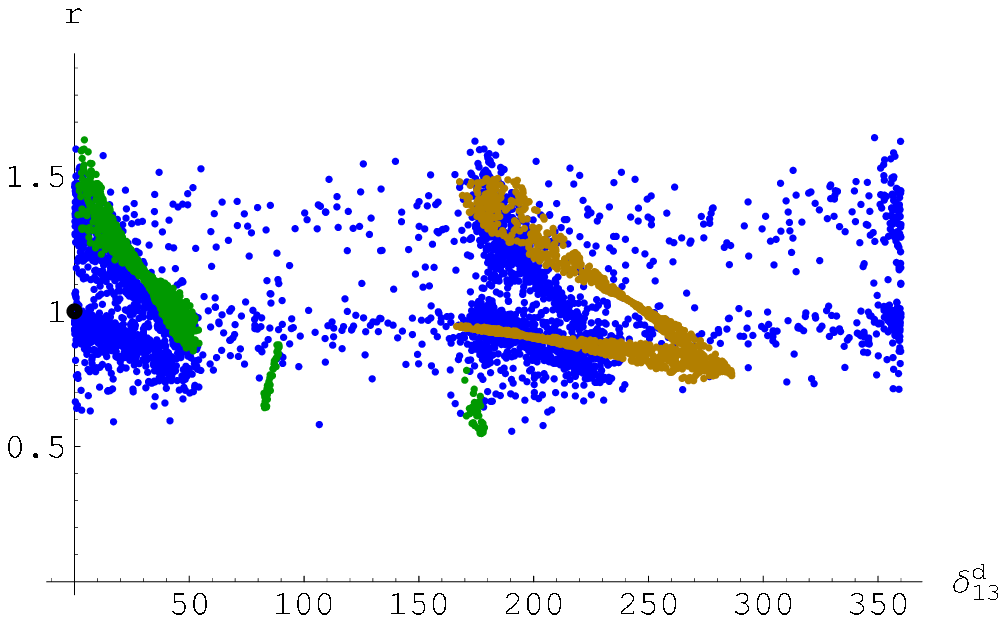,scale=.6}}
\end{minipage}
\begin{minipage}{6.3cm}
b)\vspace{-.1cm}
\center{\epsfig{file=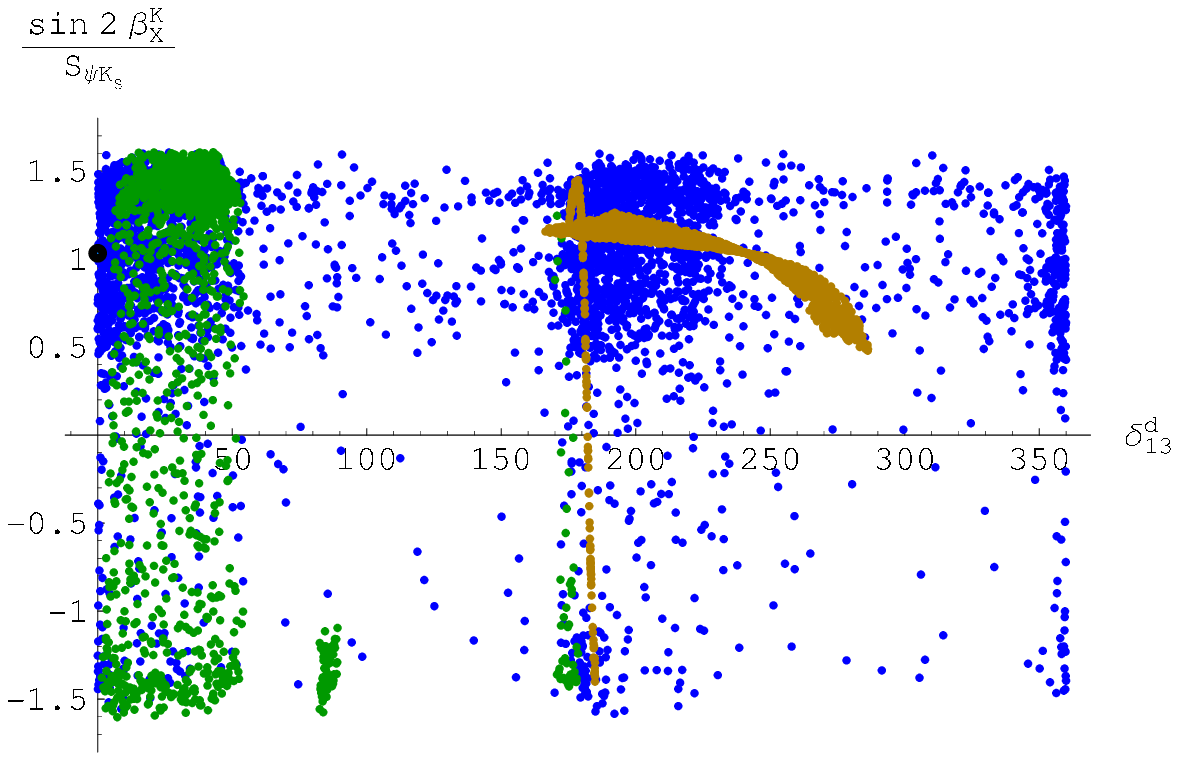,scale=.5}}
\end{minipage}
\vspace{.1cm}
\caption{The ratio $r$ of \eqref{eq:r} (a) and $\sin2\beta^K_X/S_{\psi K_S}$ (b) as functions of the new phase $\delta_{13}^d$ \cite{BBPRTUW}. The CMFV values are indicated by the black point.\label{fig:beta-ratio}}
\end{figure}

{On the other hand, sizable departures from MFV relations between $\Delta M_{s,d}$ and
    $Br(B_{s,d} \to \mu^+ \mu^-)$ and between $S_{\psi K_S}$ and the $K \to
    \pi \nu \bar \nu$ decay rates are possible.} Due to the breakdown of universality, the CMFV relation between $B_{d,s}\to\mu^+\mu^-$ and $\Delta M_{d,s}$ \cite{Buras-rel} gets modified as follows:
 \be\label{eq:r}
\frac{Br(B_s\to\mu^+\mu^-)}{Br(B_d\to\mu^+\mu^-)}= \frac{\hat
B_{B_d}}{\hat B_{B_s}} \frac{\tau(B_s)}{\tau(B_d)} \frac{\Delta
M_s}{\Delta M_d}\,r\,,\quad r= \left|\frac{Y_s}{Y_d}\right|^2
\frac{C_{B_d}}{C_{B_s}}\,,
\ee
where $C_{B_q}=\Delta M_q/(\Delta M_q)_\text{SM}$ and $Y_{q}$ are the generalizations of the known SM $Y$ function to the LHT model. Already in this CP-conserving case, the deviation of $r$ from its CMFV value $r = 1$ can amount to $50\%$, as shown in Fig.~\ref{fig:beta-ratio}.a). Even larger deviations are found, see Fig.~\ref{fig:beta-ratio}.b), for the ratio of $\sin 2\beta_X^K$, entering the $K\to\pi\nu\bar\nu$ decays, over $\sin(2\beta+2\varphi_{B_d})=S_{\psi K_S}$, which is equal to unity in MFV models \cite{beta-rel}. As $S_{\psi K_S}$ is in rough agreement with its SM prediction, a large new phase, necessary to explain the sizable deviations from MFV found in the LHT model, can only appear in the $K\to\pi\nu\bar\nu$ system. {Similar departures from MFV have very recently been found and discussed in a $Z'$ model in \cite{PSS}, where references to earlier literature can be found.}

To summarize, the largest possible effects of mirror quarks are found in rare $K$ decays and in CP-violating observables in the $B_s$ system. In particular, even if all observables related to $B_{d,s}$ physics turn out to be SM-like, large departures from the SM in $K$ physics, in particular the $K\to\pi\nu\bar\nu$ and $K_L\to\pi^0\ell^+\ell^-$ decays, would  still be possible. Therefore measurements of these decays are strongly desired.

\newsection{Charged Lepton Flavour Violation}

\subsection{Experimental Status and Prospects}

Presently, the most stringent upper bounds on {charged LFV processes} are available for $\mu\to e$ transitions, namely
\begin{gather}
Br(\mu\to e\gamma)<1.2\cdot 10^{-11}\text{\;\;\cite{muegamma}}\,,\qquad Br(\mu\to eee)<1.0\cdot 10^{-12}\text{\;\;\cite{meee}}\,,\\
R(\mu\text{Ti}\to e\text{Ti})<4.3\cdot 10^{-12}\text{\;\;\cite{mue-conv_bound}}\,,
\end{gather}
where $R(\mu\text{Ti}\to e\text{Ti})$ denotes the $\mu-e$ conversion rate in Ti. Already within the next two years, the MEG experiment at PSI \cite{megexp}
should be able to test $Br(\mu\to e\gamma)$ at the level of
$\ord(10^{-13}-10^{-14})$. Very important will also be an improved upper bound on $\mu-e$ conversion in Ti, for which the dedicated J-PARC experiment PRISM/PRIME \cite{J-PARK} should reach the sensitivity of $\ord(10^{-18})$.

On the other hand, improved experimental upper bounds on semi-leptonic and radiative $\tau$ decays have recently been presented {\cite{Belle,BaBar,Banerjee}} and further improvements are expected by the Super Flavour Factory \cite{SuperB} and SuperKEKB.

Finally let us mention that a stringent upper bound exists also for the decay $K_L\to\mu e$, that is flavour violating both in the quark and lepton sector. It reads \cite{KLmue-exp}
\be
Br(K_L\to\mu e)<4.7\cdot10^{-12} \,.
\ee

\subsection{LFV in the LHT Model}

LFV processes in the LHT model have for the first time been discussed in \cite{IndianLFV}, where the decays $\ell_i\to\ell_j\gamma$ and $\tau\to\mu\pi$ have been considered. Further, the new contributions to $(g-2)_\mu$ in the LHT model have been calculated by these authors. In \cite{BBDPT} the analysis of LFV in the LHT model has been considerably extended, and includes the decays $\ell_i\to\ell_j\gamma$, $\mu\to eee$, the six three body leptonic decays $\tau^-\to\ell_i^-\ell_j^+\ell_k^-$, the semi-leptonic decays $\tau\to\ell\pi,\ell\eta,\ell\eta'$ and the decays $K_{L,S}\to\mu e$, $K_{L,S}\to\pi^0\mu e$ and $B_{d,s}\to\ell_i\ell_j$ that are flavour violating both in the {quark and lepton sector}. Moreover, $\mu-e$ conversion in nuclei and the flavour conserving $(g-2)_\mu$ have been studied. Furthermore, a detailed phenomenological analysis has been performed in that paper, paying particular attention to various ratios of LFV branching ratios that will be useful for a clear distinction of the LHT model from the MSSM.

In contrast to $K$ and $B$ physics in the LHT model, where the SM {contributions constitute} a sizable and often the  dominant part, the T-even contributions to LFV observables are completely negligible due to the smallness of neutrino masses and the LFV decays considered are entirely governed by mirror fermion contributions. 

\begin{figure}
\begin{minipage}{6.1cm}
a)\vspace{-.4cm}
\center{\epsfig{file=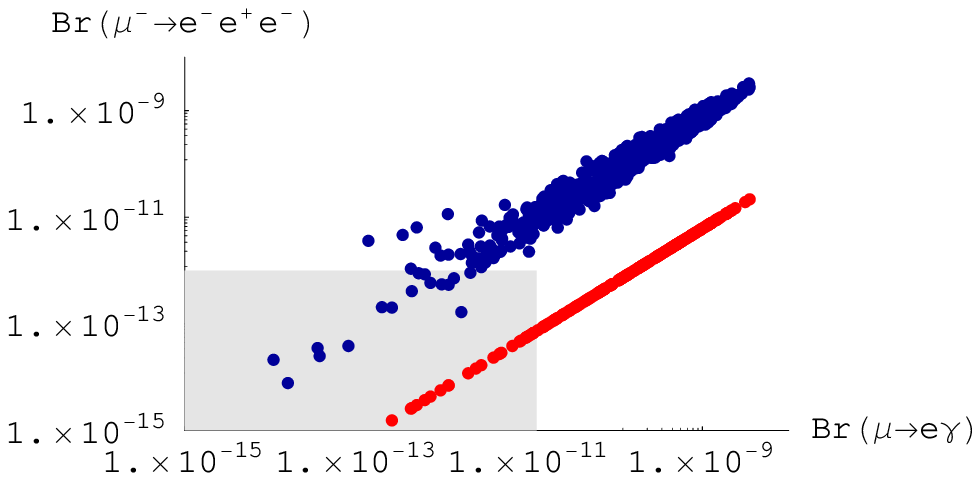,scale=0.615}}
\end{minipage}
\begin{minipage}{6.3cm}
b)\vspace{-.4cm}
\center{\epsfig{file=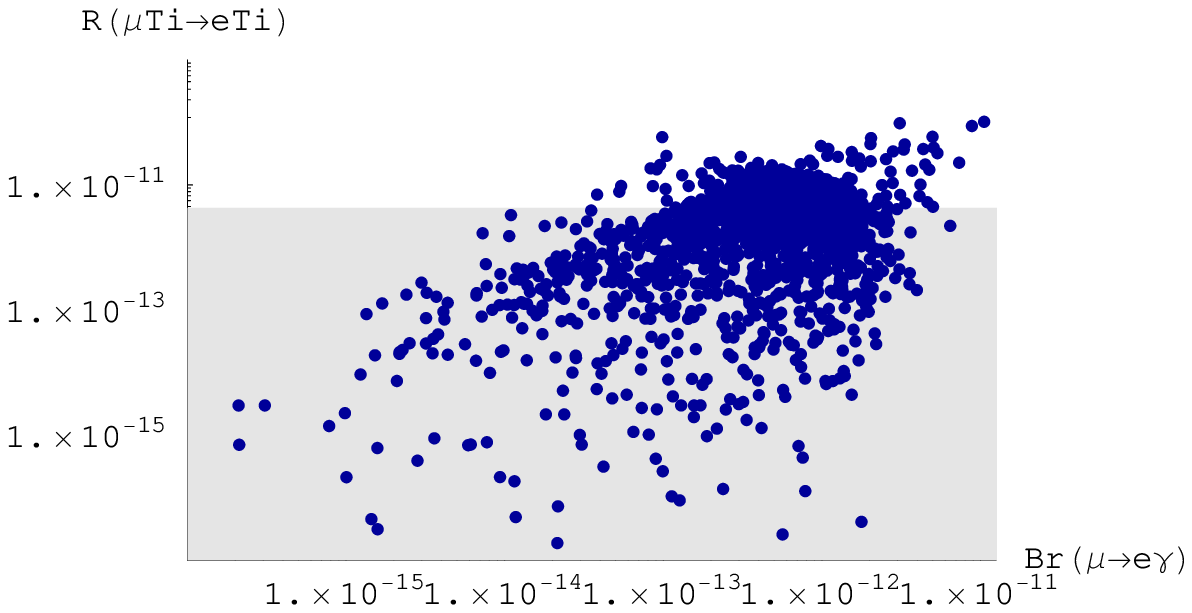,scale=0.5133}}
\end{minipage}
\vspace{-.1cm}
\caption{a) Correlation between $Br(\mu\to e\gamma)$ and $Br(\mu\to eee)$ in the LHT model (upper curve). The lower line represents the dipole contribution to $\mu\to eee$ separately, which is the dominant contribution in the MSSM  \cite{BBDPT}. b) $R(\mu\text{Ti}\to e\text{Ti})$ as a function of $Br(\mu\to e\gamma)$, after imposing the existing constraints on $\mu\to e\gamma$ and $\mu\to eee$ \cite{BBDPT}. The present experimental upper bounds are indicated by the shadowed areas. \label{fig:meg-m3e}}
\end{figure}

In order to see how large these contributions can possibly be, it is useful to consider first those decays for which the strongest constraints exist. Therefore Fig.~\ref{fig:meg-m3e}.a) shows $Br(\mu\to eee)$ as a function of $Br(\mu\to e\gamma)$, obtained from a general scan over the mirror lepton parameter space, with $f=1\tev$. {It is found that} in order to fulfil the present bounds, either the mirror lepton spectrum has to be quasi-degenerate or the $V_{H\ell}$ matrix must be very hierarchical. Moreover, as shown in Fig.~\ref{fig:meg-m3e}.b), even after imposing the constraints on $\mu\to e\gamma$ and $\mu\to eee$, the $\mu-e$ conversion rate in Ti is very likely to be found close to its current bound, and for some regions of the mirror lepton parameter space even violates this bound.

\begin{table}
{
\begin{center}
\begin{tabular}{|c|c|c|c|}
\hline
decay & $f=1000\gev$ & $f=500\gev$ & exp.~upper bound \\\hline
$\tau\to e\gamma$ & $8\cdot 10^{-10}$  & $1\cdot 10^{-8}$  & $9.4\cdot10^{-8}$ \cite{Banerjee} \\
$\tau\to \mu\gamma$ & $8\cdot 10^{-10}$ &$2\cdot 10^{-8}$  &$1.6\cdot10^{-8}$ \cite{Banerjee}\\
$\tau^-\to e^-e^+e^-$ & $7\cdot10^{-10}$ & $3\cdot10^{-8}$ & $2.0\cdot10^{-7}$ \cite{AUBERT04J}\\
$\tau^-\to \mu^-\mu^+\mu^-$ & $7\cdot10^{-10}$ & $3\cdot10^{-8}$   & $1.9\cdot10^{-7}$ \cite{AUBERT04J} \\
$\tau^-\to e^-\mu^+\mu^-$ & $5\cdot10^{-10}$ & $2\cdot10^{-8}$   & $2.0\cdot10^{-7}$ \cite{YUSA04}\\
$\tau^-\to \mu^-e^+e^-$ & $5\cdot10^{-10}$ & $2\cdot10^{-8}$  &$1.9\cdot10^{-7}$ \cite{YUSA04} \\
$\tau^-\to \mu^-e^+\mu^-$ & $5\cdot10^{-14}$  & $5\cdot10^{-14}$ & $1.3\cdot10^{-7}$ \cite{AUBERT04J}\\
$\tau^-\to e^-\mu^+e^-$ & $5\cdot10^{-14}$  &$4\cdot10^{-14}$   & $1.1\cdot10^{-7}$ \cite{AUBERT04J} \\
$\tau\to\mu\pi$ & $2\cdot10^{-9} $  & $5.8\cdot10^{-8} $ & $5.8\cdot10^{-8}$ \cite{Banerjee}\\
$\tau\to e\pi$ & $2\cdot10^{-9} $ & $4.4\cdot10^{-8} $& $4.4\cdot10^{-8}$ \cite{Banerjee}\\
$\tau\to\mu\eta$ & $6\cdot10^{-10}$ & $2\cdot10^{-8}$  &  $5.1\cdot 10^{-8}$ \cite{Banerjee}\\
$\tau\to e\eta$ & $6\cdot10^{-10}$  & $2\cdot10^{-8}$  &  $4.5\cdot 10^{-8}$ \cite{Banerjee}\\
$\tau\to \mu\eta'$ & $7\cdot10^{-10}$ & $3\cdot10^{-8}$ & $5.3\cdot 10^{-8}$ \cite{Banerjee}\\
$\tau\to e\eta'$ & $7\cdot10^{-10}$ & $3\cdot10^{-8}$  & $9.0\cdot 10^{-8}$ \cite{Banerjee}\\\hline
\end{tabular}
\end{center}
}
\caption{Upper bounds on LFV $\tau$ decay branching ratios in the LHT model, for two different values of the scale $f$, after imposing the constraints on $\mu\to e\gamma$ and $\mu\to eee$ \cite{BBDPT}. For $f=500\gev$, also the bounds on $\tau\to\mu\pi,e\pi$ have been included. The current experimental upper bounds are also given. {The bounds in \cite{Banerjee} have been obtained by combining Belle \cite{Belle} and BaBar \cite{BaBar} results.} \label{tab:bounds}}
\end{table}

The existing constraints on LFV $\tau$ decays are still relatively weak, so that they presently do not provide a useful constraint on the LHT parameter space. However, as seen in Table~\ref{tab:bounds}, most branching ratios in the LHT model can reach the present experimental upper bounds, in particular for low values of $f$, and are very interesting in view of new experiments taking place in this and the coming decade. 

The situation is different in the case of $K_L\to\mu e$, $K_L\to\pi^0\mu e$ and $B_{d,s}\to\ell_i\ell_k$, due to the double GIM suppression in the quark and lepton sectors. E.\,g. $Br(K_L\to\mu e)$ can reach values of at most $3\cdot10^{-13}$ which is still one order of magnitude below the current bound, and  $K_L\to\pi^0\mu e$ is even by two orders of magnitude smaller. Still, measuring the rates for  $K_L\to\mu e$ and $K_L\to\pi^0\mu e$ would be desirable, as, due to their sensitivity to $\RE (\xi_i^{(K)})$ and $\IM (\xi_i^{(K)})$ respectively, these decays can shed light on the complex phases present in the mirror quark sector.

While the possible huge enhancements of LFV branching ratios in the LHT model are clearly interesting, such effects are common to many other NP models, such as the MSSM, and therefore cannot be used to distinguish these models. However, correlations between various branching ratios should allow a clear distinction of the LHT model from
the MSSM. While in the MSSM \cite{MSSMrel,MSSMHiggs} the very often dominant role in decays with three leptons in the final state and in $\mu-e$ conversion in nuclei is played by the dipole operator, in \cite{BBDPT} it is found that this operator is basically irrelevant in the LHT model, where
$Z^0$-penguin and box diagram contributions are {much more important}. As can be seen in Table \ref{tab:ratios} and also in Fig.~\ref{fig:meg-m3e}.a) this implies a striking difference of various ratios of branching ratios in the MSSM and in the LHT model and should be very useful in distinguishing these two models. Even if for some decays this distinction is less clear when significant Higgs contributions are present \cite{MSSMHiggs}, it should be easier than through high-energy processes at LHC.

\begin{table}
{\renewcommand{\arraystretch}{1.5}
\begin{center}
\begin{tabular}{|c|c|c|c|}
\hline
ratio & LHT  & MSSM (dipole) & MSSM (Higgs) \\\hline
$\frac{Br(\mu^-\to e^-e^+e^-)}{Br(\mu\to e\gamma)}$  & \hspace{.6cm} 0.4\dots2.5\hspace{.6cm}  & $\sim6\cdot10^{-3}$ &$\sim6\cdot10^{-3}$  \\
$\frac{Br(\tau^-\to e^-e^+e^-)}{Br(\tau\to e\gamma)}$   & 0.4\dots2.3     &$\sim1\cdot10^{-2}$ & ${\sim1\cdot10^{-2}}$\\
$\frac{Br(\tau^-\to \mu^-\mu^+\mu^-)}{Br(\tau\to \mu\gamma)}$  &0.4\dots2.3     &$\sim2\cdot10^{-3}$ & $0.06\dots0.1$ \\
$\frac{Br(\tau^-\to e^-\mu^+\mu^-)}{Br(\tau\to e\gamma)}$  & 0.3\dots1.6     &$\sim2\cdot10^{-3}$ & $0.02\dots0.04$ \\
$\frac{Br(\tau^-\to \mu^-e^+e^-)}{Br(\tau\to \mu\gamma)}$  & 0.3\dots1.6    &$\sim1\cdot10^{-2}$ & ${\sim1\cdot10^{-2}}$\\
$\frac{Br(\tau^-\to e^-e^+e^-)}{Br(\tau^-\to e^-\mu^+\mu^-)}$     & 1.3\dots1.7   &$\sim5$ & 0.3\dots0.5\\
$\frac{Br(\tau^-\to \mu^-\mu^+\mu^-)}{Br(\tau^-\to \mu^-e^+e^-)}$   & 1.2\dots1.6    &$\sim0.2$ & 5\dots10 \\
$\frac{R(\mu\text{Ti}\to e\text{Ti})}{Br(\mu\to e\gamma)}$  & $10^{-2}\dots 10^2$     & $\sim 5\cdot 10^{-3}$ & $0.08\dots0.15$ \\\hline
\end{tabular}
\end{center}\renewcommand{\arraystretch}{1.0}
}
\caption{Comparison of various ratios of branching ratios in the LHT model and in the MSSM without and with significant Higgs contributions \cite{BBDPT}.\label{tab:ratios}}
\end{table}

Another possibility to distinguish different NP models 
through LFV processes is given by the measurement of $\mu\to e\gamma$ with polarized muons. Measuring the angular distribution of the outgoing electrons, one can determine the size of left- and right-handed contributions separately \cite{polarized-meg}. In addition, detecting also the electron spin would yield information on the relative phase between these two contributions \cite{Farzan}. We recall that the LHT model is peculiar in this respect as it does not involve any right-handed contribution.

On the other hand, the contribution of mirror leptons to $(g-2)_\mu$, being a flavour conserving observable, is negligible \cite{IndianLFV,BBDPT}. {This should also be contrasted with the MSSM with large $\tan\beta$ and not too heavy scalars, where those corrections could be significant, thus allowing to solve the possible discrepancy between SM prediction and experimental data \cite{EWg-2}.}

\newsection{Conclusions}

The LHT model provides a picture of FCNC processes at scales above the electroweak scale that differs dramatically from the SM one. While the departures from the SM expectations in the quark sector can be very large, truly spectacular effects of NP can be seen in LFV processes. Our short guide to the recent progress on FCNC processes in the LHT model will hopefully motivate the readers to have a closer look at this fascinating subject. 

We emphasize that various correlations among rare decays, both in the quark and lepton sector, will help to distinguish the LHT picture of short distance physics from other pictures, in particular from the supersymmetric one.

{\bf Acknowledgements}

AJB would like to thank the organizers for the wonderful hospitality  during the workshop in Kazimierz. 
Special thanks go to the co-authors of the work presented here: Bj{\"o}rn Duling, Anton Poschenrieder, Stefan Recksiegel, Cecilia Tarantino, Selma Uhlig and Andreas Weiler. We also thank Cecilia Tarantino for a very careful reading of the manuscript. This work has partially been supported by the Cluster of Excellence `Origin and Structure of the Universe' and by the German `Bundesministerium f{\"u}r Bildung und
Forschung' under contract 05HT6WOA.

\vfill\eject

\end{document}